\documentclass[twocolumn]{aastex631}
\usepackage{amsmath}
\usepackage{CJKutf8}

\usepackage{threeparttable}

\begin{document}
\title{Detecting Planetary Oblateness in the Era of JWST: A Case Study of Kepler-167e}

\correspondingauthor{Wei Zhu}
\email{weizhu@tsinghua.edu.cn}

\author[0009-0007-6412-0545]{Quanyi Liu\begin{CJK*}{UTF8}{gbsn}（刘权毅）\end{CJK*}}
\affiliation{Department of Astronomy, Tsinghua University, Beijing 100084, China}
\affiliation{School of Physics, Nankai University, Tianjin 300350, China}

\author[0000-0003-4027-4711]{Wei Zhu\begin{CJK*}{UTF8}{gbsn}（祝伟）\end{CJK*}}
\affiliation{Department of Astronomy, Tsinghua University, Beijing 100084, China}

\author[0000-0003-2969-6040]{Yifan Zhou\begin{CJK*}{UTF8}{gbsn}（周一凡）\end{CJK*}}
\affiliation{Department of Astronomy, University of Virginia, 530 McCormick Rd, Charlottesville, VA 22904, USA}

\author[0009-0000-6461-5256]{Zhecheng Hu\begin{CJK*}{UTF8}{gbsn}（胡哲程）\end{CJK*}}
\affiliation{Department of Astronomy, Tsinghua University, Beijing 100084, China}

\author[0000-0001-5695-8734]{Zitao Lin\begin{CJK*}{UTF8}{gbsn}（林子滔）\end{CJK*}}
\affiliation{Department of Astronomy, Tsinghua University, Beijing 100084, China}

\author[0000-0002-8958-0683]{Fei Dai\begin{CJK*}{UTF8}{gbsn}（戴飞）\end{CJK*}}
\affiliation{Institute for Astronomy, University of Hawaii, 2680 Woodlawn Drive, Honolulu, HI 96822, USA}

\author[0000-0003-1298-9699]{Kento Masuda\begin{CJK*}{UTF8}{min}（増田賢人）\end{CJK*}}
\affiliation{Department of Earth and Space Science, Graduate School of Science, Osaka University, 1-1 machikaneyama, Toyonaka, Osaka 560-0043, Japan}

\author[0000-0002-6937-9034]{Sharon~X.~Wang\begin{CJK*}{UTF8}{gbsn}（王雪凇）\end{CJK*}}
\affiliation{Department of Astronomy, Tsinghua University, Beijing 100084, China}



\begin{abstract}
Planets may be rotationally flattened, and their oblateness thus provide useful information on their formation and evolution.
Here we develop a new algorithm that can compute the transit light curve due to an oblate planet very efficiently and use it to study the detectability of planet oblateness (and spin obliquity) with the James Webb Space Telescope (JWST).
Using the Jupiter analog, Kepler-167e, as an example, we show that observations of a single transit with JWST are able to detect a Saturn-like oblateness ($f=0.1$) with high confidence, or set a stringent upper limit on the oblateness parameter, as long as the planetary spin is slightly misaligned ($\gtrsim 20^\circ$) with respect to its orbital direction. Based on known obliquity measurements and theoretical arguments, it is reasonable to believe that this level of misalignment may be common.
We estimate the sensitivity limit of JWST in oblateness detections and highlight the importance of better characterizations of cold planets in planning future JWST transit observations. The potential to detect rings, moons, and atmospheric species of the cold giants with JWST is also discussed.

\end{abstract}

\keywords{Exoplanets (498) --- Transit photometry (1709) --- Oblateness (1143) --- James Webb Space Telescope (2291)}


\section{Introduction} \label{sec:intro}

\begin{figure*}
    \centering
    \includegraphics[width=0.85\textwidth]{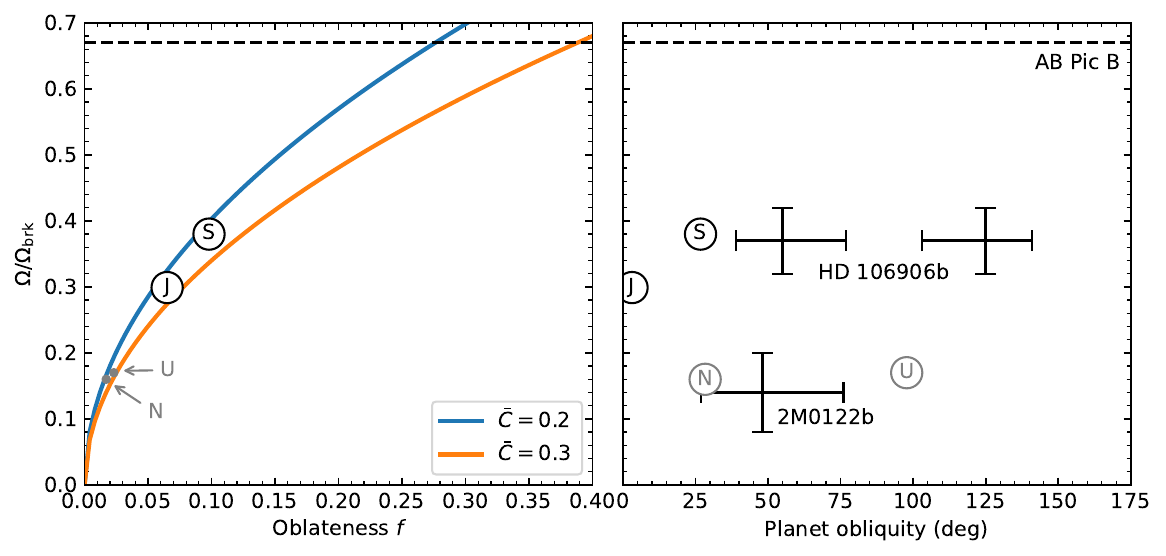}
    \caption{\textit{Left panel}: Solar system giants with spin rates and oblateness measurements. The Darwin--Radau relation (Equation~\ref{eqn:darwin-radau}) with a scaled moment of inertia $\bar{C}$ in the range $0.2$--$0.3$ provides a good estimation of the spin--oblateness relation for the solar giants. \textit{Right panel}: Objects with planet obliquity and spin rate measurements, including the solar system giants and two extra-solar planetary-mass objects (PMOs). The spin rate measurements of HD 106906b and 2M0122b are from \citet{Zhou:2020} and \citet{Zhou:2019}, respectively, and the 3D true obliquity of HD 106906b (with two degenerate solutions) and the projected obliquity of 2M0122b are from \citet{Bryan:2021} and \citet{Bryan:2020}, respectively. In terms of the fraction of break-up rotation, AB Pic B is so far the fastest rotating PMO \citep[assuming the light curve variation is indeed due to rotational modulation,][]{Zhou:2019}. Compared to Jupiter, Saturn, and the two extra-solar PMOs, Uranus and Neptune are less massive and have different formation and evolution pathways. These two objects are therefore marked in gray.} 
    \label{fig:known-systems}
\end{figure*}

Giant planets spin fast. The orbital periods of Jupiter and Saturn are both around 10\,hr. In terms of the break-up spin rate, which is defined as $\Omega_{\rm brk} \equiv \sqrt{GM_{\rm p}/R_{\rm p}^3}$, with $M_{\rm p}$ and $R_{\rm p}$ the planetary mass and (mean) radius, respectively, Jupiter and Saturn are currently spinning at about 30--40\% of the break-up spin rate. The two ice giants, Uranus and Neptune, spin relatively slower but still at about 16\% of their break-up rates.
There have also been spin period measurements on a handful of extra-solar planetary-mass objects (PMOs), suggesting that these young objects are rotating at $>10\%$ of their break-up rates (e.g., \citealt{Snellen:2014, Zhou:2016, Zhou:2019}; see \citealt{Bryan:2020b} for a short summary). Of these, AB Pic b is the record holder and rotates at $\sim67\%$ of its break-up rotation rate, if the marginally detected variability signal is indeed due to its first-order rotational modulation \citep{Zhou:2019}. If angular momentum is conserved, these objects are expected to spin even faster as they age and shrink. In addition, a significant fraction of the brown dwarfs (BDs) in the field are also rotating fast, with the known fastest-rotating BDs spinning at $\sim40\%$ of their break-up rate \citep{Tannock:2021}. Figure~\ref{fig:known-systems} illustrates existing measurements.

Fast rotations are also expected from planet formation and evolution models. If the planet is not highly magnetized and the majority of the mass is accreted at the boundary layer, the final product is a fast-rotating object very close to the break-up rate \citep[e.g.,][]{Dong:2021, Fu:2023}. Even when the magnetic field is exceedingly strong and magnetospheric accretion takes over \citep[e.g.,][]{Zhu:2015}, the spin rate immediately after formation can still be as high as 10--30\% of the break-up rate \citep[e.g.,][]{Batygin:2018, Ginzburg:2020}. Later on, a substantial fraction of the giant planets may encounter giant impacts with less massive or even equally massive objects, resulting in fast-spinning planets with somewhat randomized planet obliquities \citep[e.g.,][]{LiLai:2020, LiLai:2021}. Other physical processes like spin--orbit resonance between planets \citep[e.g.,][]{Ward:2004, Hamilton:2004} and planet--disk interactions \citep[e.g.,][]{Millholland:2019} may also induce considerable obliquities.
Non-zero obliquities can also be produced in gravito-turbulent disks \citep{Jennings:2021}.
Planetary spin and obliquity may also be altered by the tidal interaction with the host star. However, for planets with orbital periods beyond tens of days, the tidal effect from the host stars to cause the planet to spin down is negligible \citep[e.g.,][]{Seager:2002}.

Fast rotation flattens planets. For example, the equatorial radii of Jupiter and Saturn are larger than their polar radii by $7\%$ and $10\%$, respectively. This rotational deformation is usually quantified by the oblateness parameter
\begin{equation}
    f \equiv \frac{R_{\rm eq} - R_{\rm pol}}{R_{\rm eq}} ,
\end{equation}
where $R_{\rm eq}$ and $R_{\rm pol}$ are the equatorial and polar radii of the planet, respectively. This oblateness parameter can be directly related to the spin rate of the planet. As shown in the left panel of Figure~\ref{fig:known-systems}, the Solar system giant planets can all be well modeled by the Darwin--Radau relation \citep[e.g.,][Chapter 4]{Murray:1999}
\begin{equation} \label{eqn:darwin-radau}
    \left( \frac{\Omega}{\Omega_{\rm brk}} \right)^2 = \left[ \frac{5}{2}\left(1-\frac{3}{2}\bar{C}\right)^2 + \frac{2}{5} \right] f .
\end{equation}
Here $\bar{C}$ is the moment of inertia of the planet in the unit of $M_{\rm p}R_{\rm p}^2$; for Solar system giants, this parameter is known to be in the range 0.2--0.3 \citep[e.g.,][]{Ward:2004, Ni:2018}.

The transit light curve produced by an oblate planet deviates from that by a perfectly spherical planet with the same cross-section \citep{Seager:2002, Hui:2002, Barnes:2003}. This deviation, which is referred to as the oblateness signal hereafter, is primarily around the ingress and egress of the transit. The maximum amplitude of the signal is 100--200\,ppm for a Jupiter-sized planet ($R_{\rm p}/R_\star=0.1$) with Saturn-like oblateness ($f=0.1$), and it scales linearly with the transit depth \citep{Zhu:2014}.

The detection of planet oblateness requires very high photometric precision that can only be achieved with space-based telescopes. Even so, previous attempts have all yielded loose upper limits or marginal/spurious detections \citep{Carter:2010a, Carter:2010b, Zhu:2014, Biersteker:2017}. Specifically, the photometric precision that can be achieved by Kepler reaches $\sim$100\,ppm on relatively bright targets only in the Long Cadence mode \citep{Borucki:2010}, and this 30\,min exposure substantially reduces the amplitude of the oblateness signal for typical Kepler planets (orbital period $P\lesssim 100\,$d).

With the launch of the James Webb Space Telescope \citep[JWST,][]{Gardner:2006, Rigby:2023}, a firm detection of the planet oblateness finally becomes possible. The advantage of JWST over previous or other existing telescopes is multi-fold. First, JWST can achieve $\lesssim$100\,ppm photometric precision with a few minutes exposures on relatively bright targets, as has been demonstrated in previous JWST transit observations \citep[e.g.,][]{Ahrer:2023, Alderson:2023, Feinstein:2023, Rustamkulov:2023}. Second, JWST orbit at L2 enables continuous coverage of long transits, which is otherwise impossible to achieve with telescopes at low-Earth orbits (e.g., HST, CHEOPS). Furthermore, JWST observing in the near-infrared (NIR) means that the stellar noise is lower and that the limb-darkening effect is weaker, thus further boosting the signal-to-noise ratio. Finally, the broad wavelength coverage and spectroscopic capability of JWST also allow for verification of the potential oblateness signal across multiple wavelength ranges.

This paper studies the potential of JWST in detecting planet oblateness through the transit method. We use as an example the transiting Jupiter analog, Kepler-167e, which is a Jupiter-mass and Jupiter-size planet with $>1000\,$-day orbital period \citep{Kipping:2016, Chachan:2022}.
The structure of this paper is as follows:
Section~\ref{sec:suitability} summarizes the properties of the oblateness signal and identifies the suitable targets, Section~\ref{sec:simulation} investigates the detectability of the oblateness signal via the injection--recovery exercise, using Kepler-167e as an example, and finally in Section~\ref{sec:discussion} we summarize and briefly discuss the results.

\section{Suitable Targets for Oblateness Detections} \label{sec:suitability}

\begin{figure}
    \centering
    \includegraphics[width=\columnwidth]{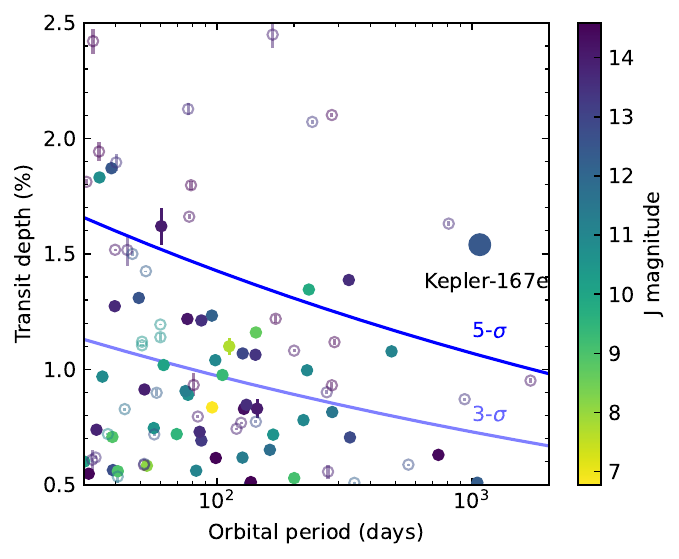}
    \caption{Confirmed (filled circles) and candidate (open circles) transiting planets in the transit depth vs.\ orbital period plane, color-coded by their $J$-band magnitudes. Data were obtained from the NASA Exoplanet Archive \citep{Akeson:2013}. Only those with relatively deep ($>0.5\%$) transits and long ($>50\,$d) periods are shown. Kepler-167e is one of the best targets for oblateness detection among all confirmed and candidate exoplanets for its long period, deep transit depth, and well-characterized properties. The blue curves indicate the 3-$\sigma$ and 5-$\sigma$ detection limits of a Saturn-like oblateness, as is calculated in Section~\ref{sec:result}.}
    \label{fig:target-selection}
\end{figure}

\begin{figure*}
    \centering
    \includegraphics[width=\textwidth]{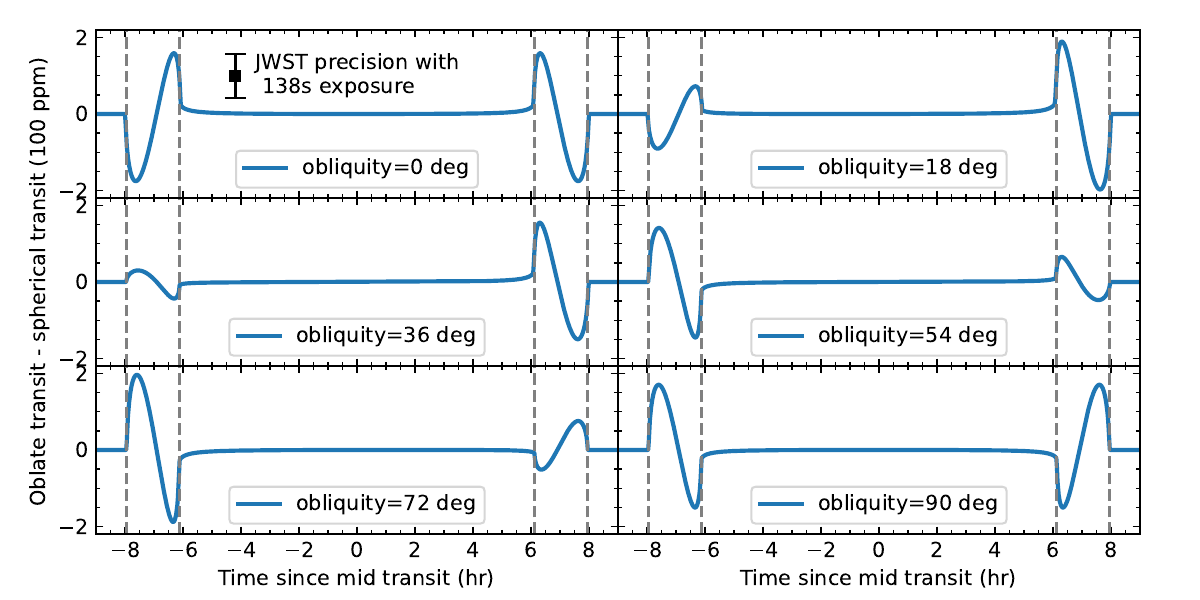}
    \caption{The oblateness signals for various planet spin obliquities. A Kepler-167e-like planet with Saturn-like oblateness ($f_\perp=0.1$) has been assumed, and a limb-darkening profile estimated for Kepler-167 in JWST NIRSpec/Prism has been adopted. The $y$-axis shows the amplitude in unit of 100\,ppm. The oblateness signal becomes asymmetric with respect to the transit center when the planet obliquity deviates from 0 or 90 deg, making it hard to be modulated by the adjustment of other parameters like orbital distance and the stellar limb-darkening profile, as our results in Section \ref{sec:result} shows. The estimated photometric precision of JWST with a 138s exposure is shown in the top left panel for a reference.}
    \label{fig:example-signals}
\end{figure*}

The transit light curve due to an oblate planet deviates from that of a perfectly spherical planet with the same cross-section primarily during the ingress and the egress. In the absence of the stellar limb-darkening effect, the maximum amplitude of the oblateness signal is given as \citep{Zhu:2014}
\begin{equation} \label{eqn:amplitude}
    \Delta \delta_{\rm max} \approx 160~{\rm{ppm}}\left(\frac{R_{\rm{p}}/R_{\star}}{0.1}\right)^2\left(\frac{f_{\perp}}{0.1}\right).
\end{equation}
Here $f_\perp \le f$ is the oblateness measured on the projected shape of the planet onto the sky plane. The inclusion of the limb-darkening effect reduces the amplitude, but this reduction is not as severe in the NIR band as it is in the optical. The actual shape of the oblateness signal also depends on other parameters, especially the projected planet obliquity $\theta_\perp$. In the case of rotational flattening, this angle measures the spin direction relative to the orbital direction. We define the positive direction to be the counterclockwise direction and set $-90^\circ \leq \theta_\perp \leq 90^\circ$. For reference, the absolute planet obliquity angles of Solar system giants and a extra-solar PMO HD 106906b \citep{Zhou:2020, Bryan:2021}, and the projected obliquity angles of 2M0122b, \citep{Zhou:2019, Bryan:2020} are shown in the right panel of Figure~\ref{fig:known-systems}. According to these known measurements, planets (or PMOs) with reasonably large obliquity angles ($\theta$ or $\theta_\perp\gtrsim 20^\circ$) are probably common. This is a useful feature for the detection of planet oblateness, as we will see in Section~\ref{sec:result}.

Planets with deep transit depths at wide orbits are therefore preferred targets for oblateness detection, because of their larger signal amplitude and presumably longer ingress and egress durations. Figure~\ref{fig:target-selection} illustrates the known confirmed and candidate transiting planets in the transit depth vs.\ orbital period plane, which were obtained from the NASA Exoplanet Archive \citep{Akeson:2013}. Among these potential targets, Kepler-167e is one of the best because of its long period, deep transit, and well-characterized properties. We therefore use it for our light curve simulation.

With an orbital period $P=1071\,$d, Kepler-167e is one of the coldest transiting exoplanets found by the Kepler mission \citep{Kipping:2016}, whose mass ($1.01\, M_J$) and radius ($0.906\,R_J$) are also very close to Jupiter \citep{Chachan:2022}. The host is a K-dwarf that is reasonably bright ($J=12.45$) and quiet, and the transit is long (16\,hr) and deep ($\sim 2\%$). Follow-up observations have also been taken to measure the mass of the planet \citep{Chachan:2022} and refine the transit ephemerides \citep{Dalba:2019, Perrocheau:2022}. The predicted mid-transit time of its next transit is BJD$_{\rm TDB}=2460609.452$ (October 25 at UT 22:49, 2024), which falls well within the JWST visibility window \citep{Perrocheau:2022}.

For illustration purposes, we show in Figure~\ref{fig:example-signals} some example oblateness signals based on the parameters of the Kepler-167 system and an assumed oblateness $f_\perp=0.1$.\footnote{One may notice that the example signals shown in Figure~\ref{fig:example-signals} are flipped compared to those in the Figure 1 of \citet{Zhu:2014}. This is because the label of $y$-axis in the latter figure was spelled incorrectly. It should be ``(Oblate planet transit) - (Spherical planet transit)'', as it is in Figure~\ref{fig:example-signals}.} Depending on the projected planet obliquity, the oblateness signal may reach an amplitude as large as $\sim 200\,$ppm and become asymmetric with respect to the mid-transit time. This is smaller than the analytic result given by Equation~(\ref{eqn:amplitude}), because here we have included the limb-darkening effect following the procedure in Section~\ref{sec:simulation}. For comparisons, the expected photometric precision of JWST on a Kepler-167-like star is $\sim 57\,$ppm for a 138\,s exposure, according to the adopted observing strategy that is described in Section~\ref{sec:simulation}.

To compute the oblateness signals shown in Figure~\ref{fig:example-signals} and, more importantly, to retrieve the model parameters through Monte Carlo sampling (Section~\ref{sec:simulation}), a fast and accurate algorithm is needed to compute the transit light curve due to an oblate planet. \citet{Carter:2010a} developed a fairly ad hoc algorithm that combined a spherical \citet{Mandel:2002} transit of a smaller-sized planet with $R_{\rm pol}$ and a Monte Carlo integration over the residual area.\footnote{Note that there are typos in their re-mapping equations. See \citep{Zhu:2014} for the correct forms.}
\citet{Rein:2019} developed a more general and more efficient algorithm called \texttt{PYPPLUSS}. By approximating the boundary as polygons, \texttt{PYPPLUSS} is able to compute the light curve of an oblate planet and/or planet with ring structures. Here we develop a new algorithm that is even faster, which we name \texttt{JoJo}.
\footnote{Named after the recently born daughter of the corresponding author.}
By applying the Green's theorem, \texttt{JoJo} turns the two-dimensional areal integral over the occultation area into a one-dimensional line integration over the boundary. As detailed in Appendix~\ref{sec:algorithm}, the final integration is reduced to a few analytic expressions and one integral that can be numerically evaluated with only dozens of points to a high precision ($<1\,$ppm). In its current form, \texttt{JoJo} works only for the quadratic limb-darkening law, which is the most widely used limb-darkening law in the literature \citep[e.g.,][]{Claret:2000, Kipping:2013a}. However, it can in principle be extended to other limb-darkening laws and transit phenomena of other types (e.g., ringed planet) after modest revisions. Compared to \texttt{PYPPLUSS}, \texttt{JoJo} is about six times faster in the case of oblate transit with Kepler-167e parameters.
The \texttt{Python} code of \texttt{JoJo} is publicly available.
\footnote{\url{https://github.com/Flippedx/JoJo}}

\section{Detectability of Planet Oblateness} \label{sec:simulation}

We perform the injection–-recovery exercise to investigate the detectability of planet oblateness (and obliquity). While the presented method is generally applicable to all suitable targets, a Kepler-167e-like transit has been assumed, meaning that the planet has a planet-to-star radius ratio $R_{\rm p}/R_\star=0.128$, an orbital period $P=1071.232\,$d, and a transit impact parameter $b=0.23$ and that the star is K-dwarf with effective temperature $T_{\rm eff}=4890\,$K, surface gravity $\log{g}=4.6$ (in cgs units), and bulk metallicity [Fe/H]$=-0.03$ \citep{Kipping:2016}.

\begin{figure}
    \centering
    \includegraphics[width=\columnwidth]{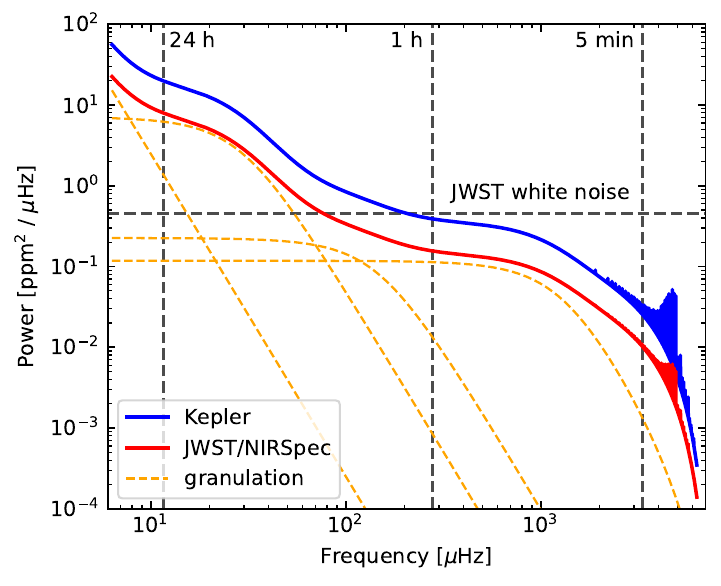}
    \caption{Theoretic stellar power spectra of a Kepler-167-like star in JWST/NIRSpec (red) and Kepler (blue) observations, generated by the \texttt{gadfly} package (Morris et al. 2024, in prep.). The multiple (super-)granulation terms in the JWST power spectrum are shown in orange dashed curves. The vertical dashed lines indicate the frequencies at certain given timescales. Within the ingress/egress duration ($1.8\,$hr) of the Kepler-167e transit, the stellar activity in JWST/NIRSpec is negligible compared to the white noise level (57\,ppm/138\,s), which is indicated by the horizontal dashed line.}
    \label{fig:psd}
\end{figure}

\begin{figure*}
    \centering
    \includegraphics[width=0.9\textwidth]{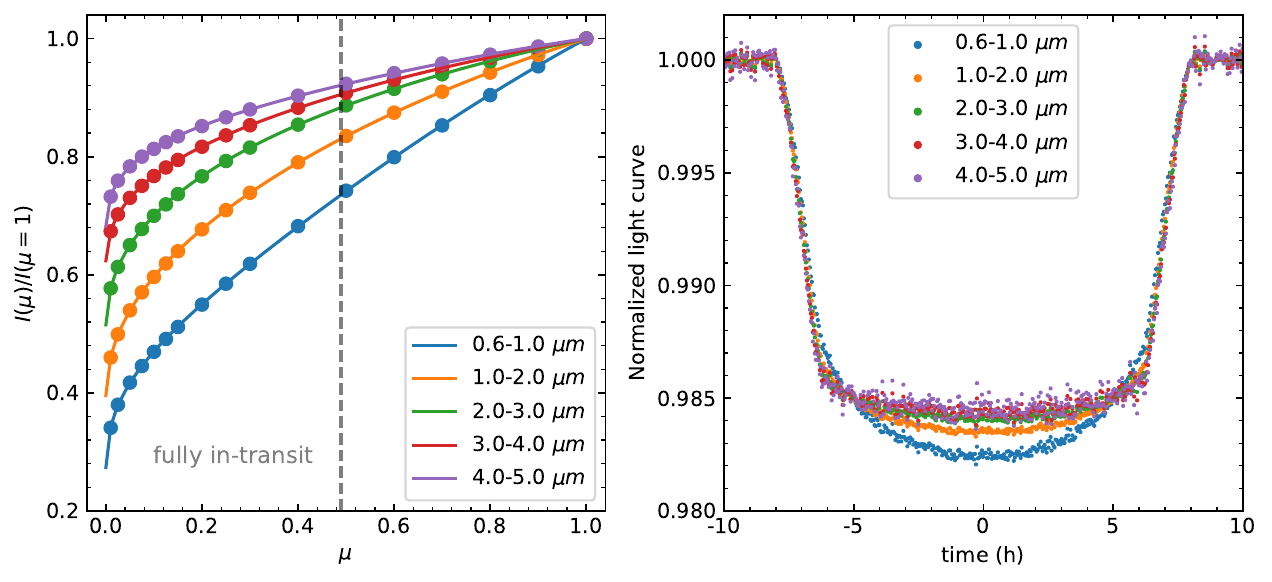}
    \caption{\textit{Left panel}: Intensity profiles across the stellar surface in different wavelength ranges of JWST/NIRSpec. Here $\mu \equiv \sqrt{1-r^2}$ and $r$ is the normalized radial coordinate on the stellar disk. The vertical dashed line denotes the location of the planet center, beyond which the planet is fully inside the stellar disk. \textit{Right panel}: Synthetic light curve data of an oblate planet transit, also divided into different wavelength ranges. Data points are sampled at the cadence of 138\,s. A Kepler-167e-like planet with a Saturn-like oblateness ($f_\perp=0.1$) and an obliquity angle of $\theta_\perp=45^\circ$ has been assumed. Note that the long-term trend has been removed for better illustration.}
    \label{fig:limb-darkening}
\end{figure*}

\subsection{JWST Light Curve Generation}\label{lc_gen}

The NIRSpec PRISM mode and the CLEAR filter are assumed for this simulation \citep{Birkmann:2022}, in order to reach high throughput and a broad wavelength coverage (0.6--5.3$\,\mu$m).
The observing strategy is optimized with JWST ETC \citep{Pontoppidan:2016}: for Kepler-167 with $J = 12.45$ and five groups/integration (at 1.38 s/integration), the central pixels reach 88\% saturation in the BOTS (sub512) mode, and the photometric precision integrated over all wavelengths is 573 ppm/integration. A total duration of 20\,hr is chosen, which leaves 2\,hr on each side of the transit. This is shorter than what is needed if one is interested in the transit depth at high precision (e.g., for the study of transmission spectrum).

Stellar activity may produce non-negligible red noises at certain timescales and thus should be taken into account in the simulation. We use the \texttt{gadfly} package
\footnote{\url{https://gadfly-astro.readthedocs.io}}
to generate the stellar power spectrum for the given stellar parameters and chosen instrument (Morris et al. 2024, in prep.). This package models the stellar activities as stochastic harmonic oscillators \citep{gp:2023}, whose amplitudes are estimated based on the asteroseismic scaling relations. The estimated stellar power spectrum for JWST/NIRSpec is shown as the red curve in Figure~\ref{fig:psd}. As the amplitudes of stellar oscillation and granulation both scale linearly with the inverse of the wavelength \citep[e.g.,][]{Kjeldsen:1995, Kjeldsen:2011}, the stellar activity becomes weaker in NIR. For the chosen observing strategy, the largest S/N is achieved at $\sim 1.5\,\mu$m, and thus the stellar noise is reduced by a factor of $\sim 3$ in JWST/NIRSpec than in Kepler. This level of red noise is below the photon noise at the timescales relevant to the oblateness detection, but nevertheless this red noise model is included in the light curve generation.

The limb-darkening effect of Kepler-167 in JWST/NIRSpec is estimated via the \texttt{ExoTic-LD}
\footnote{\url{https://exotic-ld.readthedocs.io}}
package \citep{Exotic}, in which the \citet{Kurucz:1993} stellar model and the JWST NIRSpec/Prism transmission curve have been used. The generated limb-darkening profiles are shown in the left panel of Figure~\ref{fig:limb-darkening} for multiple wavelength intervals within the NIRSpec wavelength range. The effect becomes less severe at longer wavelengths, especially toward the limb of the stellar disk, which confirms that NIR is indeed more suitable for the detection of planet oblateness. The simulated profiles are then fitted with the four-parameter limb-darkening laws to obtain the limb-darkening coefficients.
we have also tried the quadratic limb-darkening laws, and the results have no significant difference compared to that given in Section~\ref{sec:result}, which suggests that the oblateness detection is robust against imperfect modelings of the limb-darkening profiles.
We also confirm that the limb-darkening coefficients produced by \texttt{ExoTic-LD} are generally consistent with those by the \texttt{ExoCTK} package \citep{ExoCTK}.

With the noise model and the limb-darkening coefficients given above, we are now able to generate the synthetic light curve data. In the default simulation with the four-parameter limb-darkening laws, the light curves are generated by \texttt{PYPPLUSS}, whereas in the test with the quadratic limb-darkening laws, the light curves are generated by \texttt{JoJo}.
For each wavelength interval, we first compute the transit light curve using the corresponding limb-darkening coefficients, and then simulate the stellar and instrumental noise based on the power spectrum (Figure~\ref{fig:psd}). The seed for the random number generator in the noise simulation has been fixed across wavelength bins, in order to produce consistent noise behaviors. We show in the right panel of Figure~\ref{fig:limb-darkening} some example light curves generated in this way. These wavelength-dependent light curve data can in principle be directly modeled, which should provide a cross-check of the oblateness signal as well as allow for the detection of atmospheric signals. However, for the purpose of this work and considering that the oblateness signal is independent of wavelength, we here combine all wavelengths and produce a single white light curve.

To investigate the detectability of oblateness, simulated light curves with different combinations of oblateness and obliquity values are generated: for oblateness, we choose $f_\perp=0.1$ (i.e., a Saturn-like oblateness) and $0$ (i.e., spherical planet); for non-zero oblateness values, we choose $\theta_\perp=0$, $20^\circ$, and $45^\circ$ as the possible planet obliquity. Because of the geometric and time reversal symmetries, light curves from these three obliquity values cover all the essential cases, even though they do not span the full dynamical range of $\theta_\perp$.

\subsection{Modeling Procedure} \label{sec:modeling}

We fit the simulated JWST light curve with both spherical and oblate transit models to determine the detectability of the injected oblateness signal. The quadratic limb-darkening law, which is different from the four-parameter law used in the injection, is used in the model fitting. This choice is on purpose, in order to test the detectability of oblateness signal in the presence of imperfect modeling of the limb-darkening effect. In the spherical model \citep{Mandel:2002}, the transit light curve is described by six parameters, namely the mid-transit time $t_0$, the impact parameter $b$, the planet-to-star radius ratio $R_{\rm p}/R_{\star}$, the stellar ``mean density'' $\rho_\star$ \citep{Seager:2003}, and the two quadratic limb-darkening coefficients $q_1$ and $q_2$ following the parameterization of \citet{Kipping:2013a}. We have fixed the orbital period given that there is only a single transit. Note that $\rho_\star$ does not necessarily equal the stellar mean density in value except for perfectly circular orbit \citep{Dawson:2012}. In the oblate model, the transit light curve contains two additional parameters, $f_\perp$ and $\theta_\perp$, and the model is computed by \texttt{JoJo}.

The simulated data contain a systematic trend that must be removed. This detrending is done via the \texttt{Wotan}
\footnote{\url{https://wotan.readthedocs.io/}}
package \citep{Wotan:2019}, and a template with the sum of sine and cosine curves is adopted. Because the out-of-transit portion is short compared to long transit duration, this detrending is done simultaneously with the transit fitting. To prevent detrending from affecting the detection of the oblateness signal, the ingress and egress regions are masked.

For a given model (spherical or oblate), the logarithm of the model likelihood is evaluated as
\begin{equation}\label{eqn:lnL}
    \ln{\mathcal{L}}=-\frac{1}{2}\sum_i\left(\frac{d_i-m_i}{e}\right)^2-\frac{n}{2}\ln(2\pi e^2) .
\end{equation}
Here $d_i$ is the simulated light curve before normalization, and $m_i \equiv f_i\cdot t_i$ is the full model including the transit model $f_i$ and the systematic trend $t_i$. The parameter $e$ accounts for the flux variation and is evaluated as the standard deviation in the residual flux $(d_i-m_i)$. Considering the non-Gaussian behavior of model parameters, especially $f_\perp$ and $\theta_\perp$ \citep{Zhu:2014}, we use the dynamic nested sampling method, implemented by the \texttt{dynesty} package \citep{Speagle:2020}, to sample the parameter space and estimate the posteriors. Generous and physically meaningful boundaries are adopted for the model parameters. In particular, we use $f_\perp\cos(2\theta_\perp)$ and $f_\perp\sin(2\theta_\perp)$ as free parameters and set their priors to be uniform distributions between $(-0.25, 0.25)$. We choose not to use the $f_\perp$ and $\theta_\perp$ directly as free parameters because $\theta_\perp$ is ill defined when $f_\perp=0$. In order to force a uniform prior on $f_\perp$ in the oblate model, we add a rectification term, $-\ln f_\perp$, in Equation~(\ref{eqn:lnL}). The iterations are stopped when the estimated contribution of the remaining prior volume to the total evidence falls below 0.01.

\subsection{Expected Results} \label{sec:result}

\begin{figure*}
    \centering
    \includegraphics[width=\textwidth]{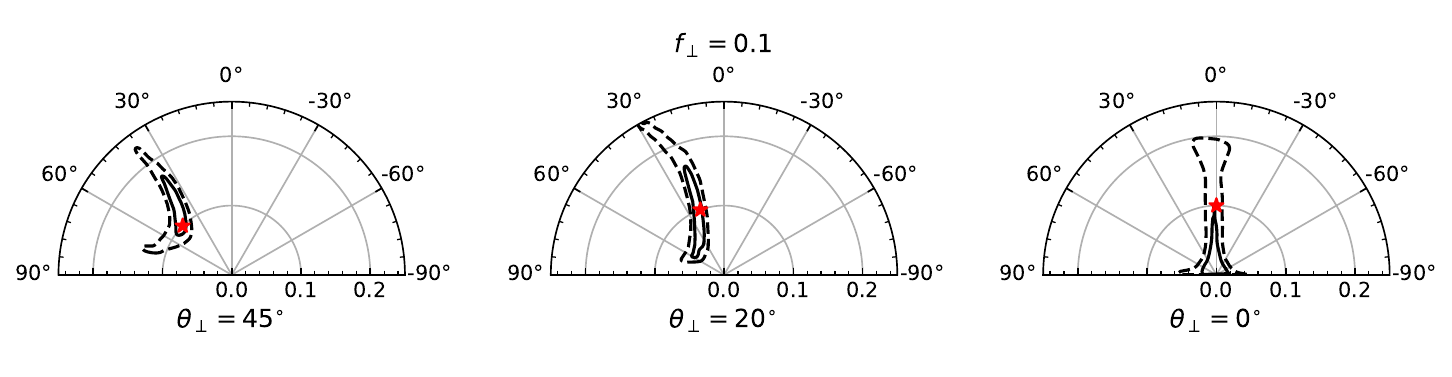}
    \caption{Constraints on the oblateness $f_\perp$ and obliquity $\theta_\perp$ in the polar coordinate from the injection--recovery exercise for $f_\perp=0.1$. Three obliquity values, namely $0^\circ$ (right panel), $20^\circ$ (middle panel), and $45^\circ$ (left panel), are chosen. Locations of the ground truth are shown as red asterisks, and the 1-$\sigma$ and 2-$\sigma$ contours from the \texttt{dynesty} modeling (Section~\ref{sec:modeling}) are projected onto the polar coordinates and shown with solid and dashed curves, respectively.}
    \label{fig:JWST-contours}
\end{figure*}

\begin{figure*}
    \centering
    \includegraphics[width=\textwidth]{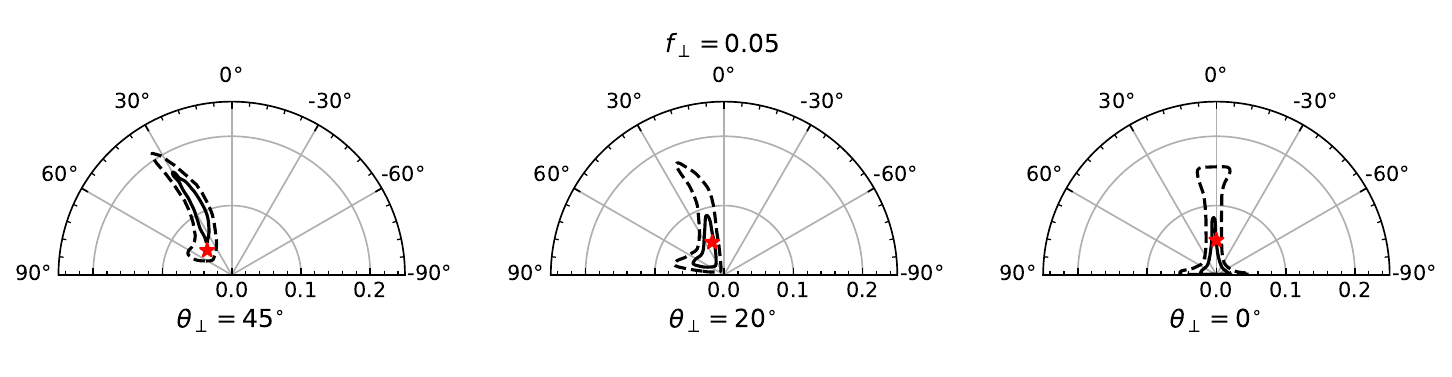}
    \caption{Similar to Figure~\ref{fig:JWST-contours} but for $f_\perp=0.05$. The detection significance of the oblateness signal is reduced with respect to the case of $f_\perp=0.1$. Nevertheless, the signal can be securely detected for $\theta_\perp=45^\circ$ and marginally detected for $\theta_\perp=20^\circ$.}
    \label{fig:JWST-contours-0.05}
\end{figure*}

\begin{figure*}
    \centering
    \includegraphics[width=0.48\linewidth]{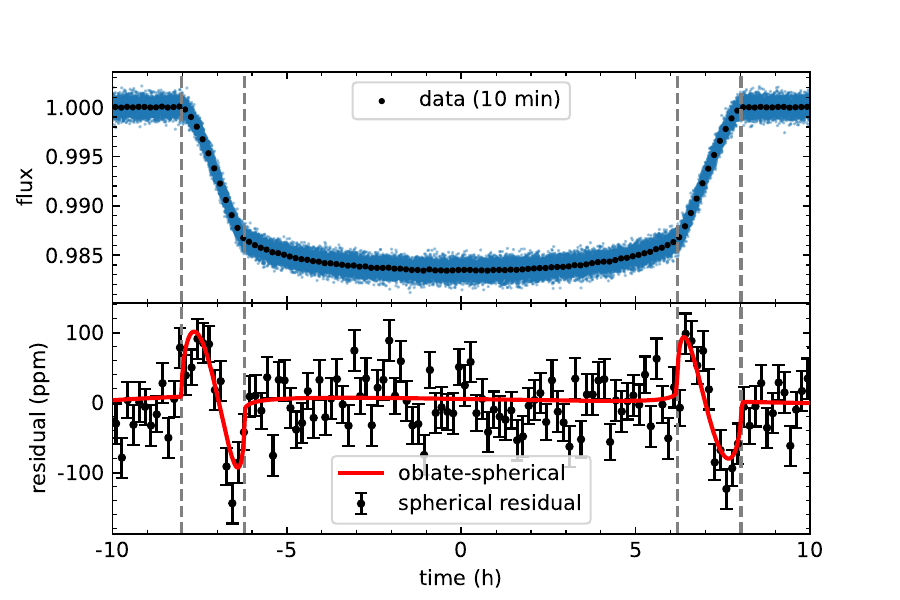}
    \includegraphics[width=0.48\linewidth]{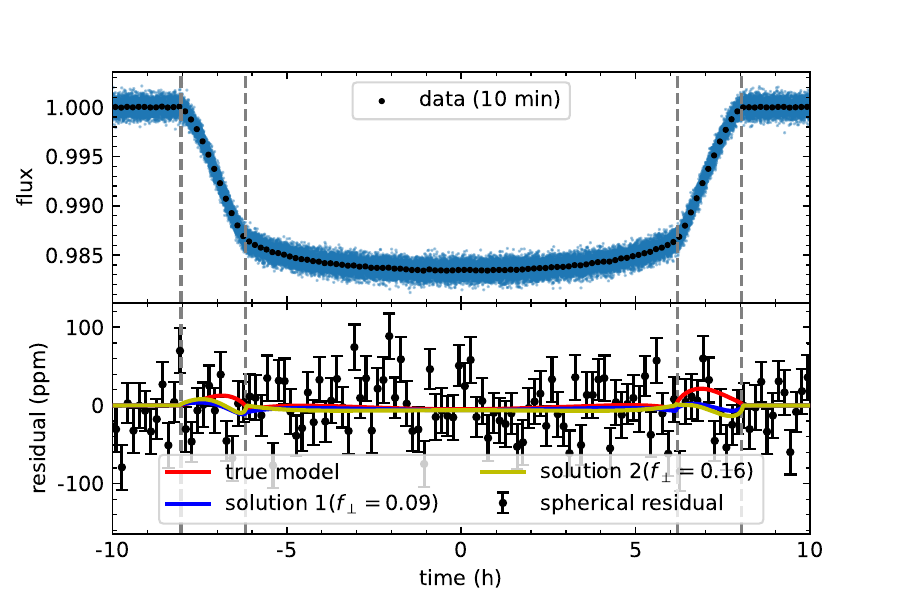}
    \caption{Simulated light curves of a Kepler-167e-like planet with Saturn-like oblateness ($f_\perp=0.1$) and the residuals of the best-fit spherical models. Different planet obliquities have been used: $\theta_\perp=45^\circ$ in the left and $\theta_\perp=0^\circ$ in the right. The raw observations with $1.38\,\rm{s}$ exposures (blue dots) are binned every 10\,min (black dots with error bars) for illustration purposes. For the case shown on the left, the spherical model cannot fit the data well during ingress and egress, and the residuals can be well modeled by the oblate model. For the case shown on the right, the light curve can be described by a spherical model and several oblate models with different oblateness values almost equally well.}
    \label{fig:JWST-res}
\end{figure*}

\begin{figure}
    \centering
    \includegraphics[width=\columnwidth]{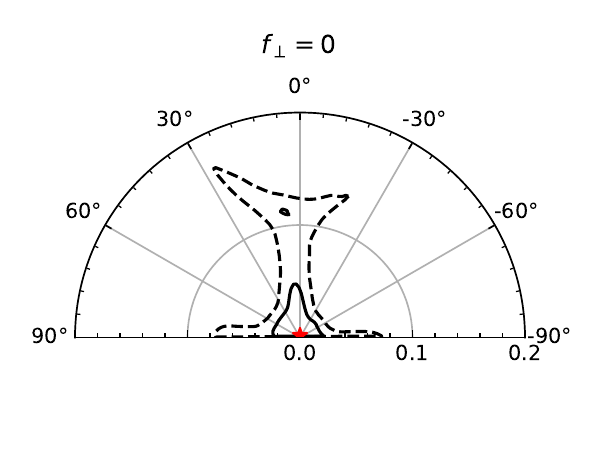}
    \caption{Similar to Figure~\ref{fig:JWST-contours} but for $f_\perp=0$.}
    \label{fig:JWST-contours-nondetection}
\end{figure}

We apply the modeling procedure to the simulated light curves and derive the posterior distributions of the model parameters, especially $f_\perp$ and $\theta_\perp$. The results are shown in Figure~\ref{fig:JWST-contours} and \ref{fig:JWST-contours-0.05} for the cases with non-zero injected oblateness values, namely $f_\perp=0.1$ and 0.05, respectively. These values correspond to a dimensionless spin rate in the range 20--40\% (see Figure~\ref{fig:known-systems}). The injected parameters can be reasonably recovered, especially the oblateness $f_\perp$, unless the planet obliquity is very close to $0^\circ$. To better demonstrate this point, we show in Figure~\ref{fig:JWST-res} the simulated data and best-fit models of two sets of oblateness configurations, $(f_\perp, \theta_\perp)=(0.1, 45^\circ)$ and $(0.1, 0^\circ)$. The former case is nearly the optimal configuration for oblateness detection in terms of the signal amplitude and especially the left--right asymmetry (e.g., Figure~\ref{fig:example-signals}). Indeed, the oblateness signal can be clearly seen in the residuals of the best-fit spherical model. In terms of the detection significance, the best-fit oblate model is favored over the best-fit spherical model by $\Delta \ln \mathcal{L}\approx 46$ at the price of two extra parameters. In contrast, the oblateness signal for the configuration with $\theta_\perp=0^\circ$ cannot be recovered, even though the theoretical amplitude of the signal is also large compared to the measurement uncertainty (Figure~\ref{fig:example-signals}). In such a configuration, the symmetric signal may be modulated by the adjustment of other parameters such as the impact parameter, orbital distance (via $\rho_\star$), and the stellar limb-darkening profiles, thus resulting in nearly flat residuals after the spherical model is subtracted from the data. As a result, the injected planet oblateness cannot be well recovered, leading to large uncertainties on the parameter $f_\perp$ along the direction of planet obliquity $\theta_\perp$.

The above simulation also allows us to estimate the detection limit for planet oblateness. With other parameters unchanged, we vary the transit depth and orbital period of the planet, until the difference in the goodness-of-fit, $\Delta \ln \mathcal{L}$, between the best-fit oblate and spherical models is reduced to $25/2$ and $9/2$, which are defined as the 5-$\sigma$ and 3-$\sigma$ limits, respectively. The resulting detection limits are shown as blue curves in Figure~\ref{fig:target-selection}. Since the nearly optimal configuration with $f_\perp=0.1$ and $\theta_\perp=45^\circ$ has been used as the injected signal, the derived limits are somewhat optimistic. Nevertheless, there are over two dozen known transiting planets (or candidates) above the derived 3-$\sigma$ sensitivity limit. With these suitable targets and potentially more from the ongoing surveys like TESS and Gaia, there will soon be a sample of exoplanets for JWST to probe the oblateness and obliquity, once its capability is demonstrated.

For completeness, we also check the possibility of a false positive detection of planet oblateness. The simulated light curve with perfectly spherical planet is modeled with the oblate model, and the resulting posterior distribution of the oblateness parameters is shown in Figure~\ref{fig:JWST-contours-nondetection}. The constraint on oblateness $f_\perp$ is again obliquity-dependent: for obliquity values close to $0^\circ$ the 2-$\sigma$ upper limit on $f_\perp$ is $>0.1$, whereas if $|\theta_\perp|\gtrsim 20^\circ$ one is able to put stringent upper limit on $f_\perp$. For the result shown in Figure~\ref{fig:JWST-contours-nondetection}, $f\lesssim 0.03$ can be obtained if the planet is misaligned by $\gtrsim 20^\circ$. For Kepler-167e, such a small oblateness would suggest that the planet spin period should be longer than $\sim11$\,hr based on Equation~\ref{eqn:darwin-radau}, thus ruling out Jupiter- and Saturn-like rotations.

A nearly spin-orbit aligned configuration reduces the capability to constrain/detect the planet oblateness substantially, as the above investigations have suggested. On the other hand, cold giants with misaligned ($\theta_\perp\gtrsim 20^\circ$) orbits may not be uncommon, as seen in Solar System giants and the extra-solar PMOs with obliquity measurements (right panel of Figure~\ref{fig:known-systems}). It is also expected from theoretical perspectives that planetary spins can often be tilted substantially via physical processes such as giant impacts \citep[e.g.,][]{LiLai:2020, LiLai:2021}, spin--orbit resonance between planets \citep[e.g.,][]{Ward:2004, Hamilton:2004}, planet--disk interactions \citep[e.g.,][]{Millholland:2019}, etc. Therefore, there is reason to believe that JWST is able to measure, or at least put physically meaningful constraints on, the planet oblateness and obliquity based on transit observations of no more than a handful of suitable targets.

\section{Summary \& Discussion} \label{sec:discussion}

This work investigates the detectability of planet oblateness and obliquity via JWST transiting observations. Our findings are summarized as following:
\begin{itemize}
    \item Planets with deep transits and long orbital periods are favorable targets for oblateness detections, because the oblateness signal occurs during ingress and egress and its amplitude scales almost linearly with the transit depth.
    \item Observations of a single transit with JWST on a Kepler-167e-like planet should be able to detect or set stringent constraints on the planet oblateness, unless the planetary spin is nearly aligned ($\lesssim 20^\circ$) with its orbital direction.
    \item Based on the obliquity measurements of Solar System giants and extra-solar PMOs as well as theoretical arguments, we believe that large spin obliquities may be common among cold extra-solar giants.
    \item Detection limits of planetary oblateness for JWST are estimated, based on which we show that the oblateness of over a dozen confirmed planets or planetary candidates may already be detectable with JWST.
\end{itemize}


The planet Kepler-167e has been chosen for this case study.
\footnote{We note that this target has been selected for JWST observations in Cycle 3 \citep{Cassese:2024}.}
This choice is made for two reasons. First, the planet has a deep transit ($\sim1.5\%$) and a long orbital period ($\sim 1071\,$d), so the potential oblateness signal is prominent. Second, both the planet and its host star are well characterized, so a realistic observing condition, wherever possible, can be imposed in the simulation. While the second reason should certainly be considered in proposing and designing JWST observations, the first reflects the physical limitation of JWST in oblateness detection. If only this first reason is considered, a few other targets may be as good as Kepler-167e, as shown in Figure~\ref{fig:target-selection}, and more may be found by the ongoing missions like TESS \citep{Ricker:2015} or possibly Gaia \citep{Perryman:2014} as well as future planned missions like ET \citep{Ge:2022} or PLATO \citep{Rauer:2014}. With the increasing number of suitable targets, more detailed and better characterizations of these long-period giants are strongly encouraged for better planning JWST observations.

While the present work focuses on the detection of planet oblateness, there are certainly other interesting science issues that can be studied with the same (or similar) dataset. Similar to the Solar System giants, giant exoplanets at relatively wide orbits may also harbour ring and moon systems \citep[e.g.,][]{Mamajek:2012, Kipping:2022}, and these structures can lead to detectable signatures in the transit light curve \citep[e.g.,][]{Barnes:2004, Kipping:2012}. 
Unlike exomoons, signals produced by exorings may appear qualitatively similar to the signals produced by oblate planets, but the former would have much larger amplitudes \citep{Barnes:2004}, and the planets would have much smaller densities if the rings are not taken into account in the light curve modeling \citep{Piro:2020}.
The search for exo-rings and exo-moons would normally require a much longer time baseline than the search for oblateness signals. With the same dataset, one can also study the atmosphere of cold giants, although with a low equilibrium temperature the atmosphere scale height is small in comparison with that of warm and hot giant planets. For Kepler-167e, if we take the model transmission spectra of \citet{Dalba:2019} and the measured planetary mass of \citet{Chachan:2022}, then the variation in transit depth is $\lesssim 50\,$ppm across the near and mid IR wavelength range. The most prominent feature is due to methane between 3--4$\,\mu$m, and yet the estimated significance is at the level of $\Delta \ln \mathcal{L}\approx 5$. We leave more detailed investigations of the bonus science topics to some future works.

\begin{acknowledgements}
    We would like to thank the anonymous referee for comments on the manuscript.
    This work is supported by the National Natural Science Foundation of China (grant No.\ 12173021 \& 12133005) and the CASSACA grant CCJRF2105. We also acknowledge the Tsinghua Astrophysics High-Performance Computing platform for providing computational and data storage resources.
\end{acknowledgements}

\appendix

\section{Use Green's theorem to compute transit light curve due to an oblate planet}\label{sec:algorithm}

The normalized flux due to a transiting object is given by
\begin{equation}
    f=1-\frac{\Delta F}{F_{\rm tot}} = 1- \frac{6 \Delta F}{(6-2u_1 - u_2)\pi} .
\end{equation}
Here we have made use of the known expression for the total flux, $F_{\rm tot}$, for a quadratic limb-darkening law in the form
\begin{equation}
    I(\mu)=1-u_1(1-\mu)-u_2(1-\mu)^2, ~ \mu\equiv\sqrt{1-(x^2+y^2)} .
\end{equation}
The Cartesian coordinates, $(x, y)$, are used to denote the position at the stellar disk relative to its center, and $u_1$ and $u_2$ are the limb-darkening parameters. The flux variation due to the transiting object, $\Delta F$, normally calculated by an areal integral over the occulted area $S$, can be transformed into a line integral over its boundary $l$
\begin{equation}
    \Delta F=\iint_S I(\mu){\rm d}s=\oint_l~(A_x{\rm d}x+A_y{\rm d}y),
\end{equation}
with $A_x$ and $A_y$ satisfying
\begin{equation}\label{Ay}
    \frac{\partial A_y}{\partial x}-\frac{\partial A_x}{\partial y}=I(\mu).
\end{equation}
For simplicity, we set $A_x = 0$ and obtain
\begin{equation} \label{eqn:Ay}
A_y = x + \frac{1}{6} \left\{ 3(u_1+2u_2) \left[(\mu-2)x + (1-y^2)\tan^{-1}\frac{x}{\mu} \right] + 2u_2 (3y^2+x^2)x \right\} ,
\end{equation}
The flux variation is then given by
\begin{equation} \label{eqn:deltaF}
\Delta F = (1-u_1-2u_2) \oint_l x{\rm d}y + \frac{u_2}{3} \oint_l (3y^2+x^2)x {\rm d}y + \frac{u_1+2u_2}{2} \oint_l \left[\mu x + (1-y^2)\tan^{-1}\frac{x}{\mu} \right] {\rm d}y .
\end{equation}
The above expression is very general and applies to the transit/occultation of various types, including oblate transit and planetary ring transit. In the oblate transit problem, the first two integrals can be expressed in closed forms once the areal edge $l$ is known. It is only the last one that requires numerical integration.

Before proceeding, let us define the coordinate system to work with. The origin is defined at the center of the stellar disk, and the $x$-axis is chosen to be the major axis of the oblate planet. The $y$-axis is chosen such that the geometric center of the oblate planet, $(x_0, y_0)$, is in the first quadrant, which can always be achieved by making use of the geometric symmetry. We will also define $d_0 \equiv \sqrt{x_0^2 + y_0^2}$ to be the distance between the centers of the planet and the star. Note that, if the planetary spin is misaligned relative to the orbital axis,  the absolute directions of both $x$ and $y$ axes may change as the transit proceeds. The coordinates are normalized by the radius of the star.

With the above definition, the stellar limb is described by a unit circle, $x^2+y^2=1$, and the coordinates of a point on the stellar limb are
\begin{equation} \label{eqn:xsys}
x_s=\cos{\beta},\quad y_s=\sin{\beta} ,
\end{equation}
where $\beta$ measures the angle from the $x$-axis. The planetary edge is described by an ellipse, $(x-x_0)^2/a^2 + (y-y_0)^2/b^2=1$, and a point on the edge of the planetary disk has coordinates
\begin{equation} \label{eqn:xpyp}
x_p=x_0+a\cos{E},\quad y_p=y_0+b\sin{E} .
\end{equation}
Here $a$ and $b$ are the semi-major and semi-minor axes of the ellipse, which approximates the projected shape of the oblate planet, respectively. The angle $E$ measures the eccentric anomaly of the point on the planetary edge.

To evaluate the integral given by Equation~(\ref{eqn:deltaF}), we need to determine the boundary $l$, and there are three different cases:
\begin{itemize}
    \item Planet fully outside the stellar disk, namely, ``out of transit.'' In this case, $\Delta F=0$ by definition.
    \item Planet fully inside the stellar disk, i.e., ``in transit.'' There is at most one intersecting point between planetary and stellar edges.
    \item The intermediate stages (``ingress'' or ``egress'') where the planetary edge intersects with the stellar edge at more than one point.
\end{itemize}
The key issue would be to identify the three cases given above, in order to treat them separately. This is achieved by solving for intersection points between the stellar and planetary edges (Appendix~\ref{sec:intersection}).

\subsection{Solving for intersection points} \label{sec:intersection}

Two special situations can be easily identified:
\begin{itemize}
    \item If $d_0 \geq (1+a)$, the planet is ``out of transit.''
    \item $d_0 \leq (1-a)$, the planet is ``in transit.''
\end{itemize}
In the general situation, the intersection points between the edges of the stellar and planetary shapes should be determined in order to evaluate the flux variation in Equation~(\ref{eqn:deltaF}). Similar to \citet{Rein:2019}, this is done by solving a quartic equation.

Using the complex coordinate $z \equiv x + iy$ and its complex conjugate $\bar{z} \equiv x - iy$, we can combine the equations describing the stellar and planetary edges into the following quartic equation
\begin{equation} \label{eqn:quartic}
c_4 z^4 + c_3 z^3 + c_2 z^2 + c_1 z + c_0 = 0 ,
\end{equation}
with the coefficients given as
\begin{equation}
\left\{ \begin{array}{rcl}
c_4 & = & a^2-b^2 \\
c_3 & = & 4b^2 x_0 - 4i a^2 y_0 \\
c_2 & = & -2 (a^2+b^2-2a^2b^2 + 2b^2x_0^2 + 2a^2y_0^2) \\
c_1 & = & 4b^2 x_0 + 4i a^2 y_0 \\
c_0 & = & a^2 - b^2
\end{array}\right..
\end{equation}
Note that in the special case with $f=0$ and thus $a=b$ (i.e., spherical planet), Equation~(\ref{eqn:quartic}) is reduced to a quadratic equation. Quartic equations can in principle be solved analytically, but these solutions may not be numerically stable. For our purpose, we use \texttt{numpy.roots} to solve Equation~(\ref{eqn:quartic}) numerically. In the limit of $R_{\rm p}/R_\star \ll 1$, there are at most two true solutions out of the four complex solutions, and they are identified based on the criterion $(x-x_0)^2/a^2+(y-y_0)^2/b^2-1<10^{-3}$. With these true solutions, we have the following different situations:
\begin{itemize}
  \item Zero or one intersection point and $d_0>1$:  out of transit.
  \item Zero or one intersection point and $d_0<1$:  in transit.
  \item Two intersection points: ingress/egress. The two intersection points are denoted as $(x_1,~y_1)$ and $(x_2,~y_2)$.
\end{itemize}

\subsection{The ``in transit'' case}

In this case, the boundary $l$ is given by the edge of the planet $l_p$, which is a full ellipse. The integral we want to compute is
\begin{equation} \label{eqn:deltaF-intransit}
\Delta F = (1-u_1-2u_2) \oint_{l_p} x_p {\rm d}y_p + \frac{u_2}{3} \oint_{l_p} (3y_p^2+x_p^2)x_p {\rm d}y_p + \frac{u_1+2u_2}{2} \oint_{l_p} \left[\mu_p x_p + (1-y_p^2)\tan^{-1}\frac{x_p}{\mu_p} \right] {\rm d}y_p .
\end{equation}
Here $\mu_p \equiv \sqrt{1-x_p^2-y_p^2}$, and $x_p$ and $y_p$ are given by Equation~(\ref{eqn:xpyp}). Of the three integrals, only the last one cannot be expressed in closed forms.
\footnote{The first term in the last integral can be expressed with elliptical integrals, but this is not very helpful given that one still needs to perform numerical integration for the second term, which is the most time-consuming.}
In the end, the flux variation is given by
\begin{equation}
\Delta F = \pi ab \left[1-u_1-2u_2 + \frac{u_2}{4}(a^2+b^2+4d_0^2)\right] + \frac{u_1+2u_2}{2} b \int_0^{2\pi} \left[\mu_p x_p + (1-y_p^2)\tan^{-1}\frac{x_p}{\mu_p}\right]\cos{E} {\rm d}E .
\end{equation}
With the parameters of Kepler-167e, the evaluation of the last integral requires 30 points in order to reach $<1$\,ppm precision.

\subsection{The ``ingress/egress'' case}

In this case the line integral is broken into two parts: $l=l_s+l_p$, with $l_s$ the segment on the stellar limb and $l_p$ the segment on the planetary edge that is inside the stellar disk. With our convention ($x_2<x_1$, $x_0\geq 0$, and $y_0\geq 0$), the integration goes from $(x_1, y_1)$ to $(x_2, y_2)$ on $l_s$ and from $(x_2, y_2)$ to $(x_1, y_1)$ on $l_p$. It can be shown that the flux variation (Equation~\ref{eqn:deltaF}) is now given by
\begin{equation} \label{eqn:limb_contribution}
\begin{aligned}
\Delta F = &\int_{l_s} \left[ (1-u_1-2u_2)x_s + \frac{u_2}{3} (2y_s^2+1)x_s \right] {\rm d}y_s + (1-u_1-2u_2)\int_{l_p} x_p {\rm d}y_p + \frac{u_2}{3} \int_{l_p} (3y_p^2+x_p^2)x_p {\rm d}y_p \\
&+ \frac{u_1+2u_2}{2} \int_{l_p} \left\{ \mu_p x_p + (1-y_p^2) \left[ \tan^{-1}\frac{x_p}{\mu_p} - \frac{\pi}{2} \mathrm{sgn}(x_p) \right] \right\} {\rm d}y_p .
\end{aligned}
\end{equation}
Here $\mathrm{sgn}(x_p)$ takes the sign of $x_p$, and we have made use of the fact that $\mu_s=0$ for points on $l_s$. The first three integrals yield
\begin{equation}
    \int_{l_s} (\cdot) {\rm d} y_s = \left(\frac{1}{2}-\frac{u_1}{2}-\frac{3}{4}u_2\right) (\beta_2 - \beta_1) +
\left[\left(\frac{1}{2}-\frac{u_1}{2}-\frac{11}{12}u_2\right)(x_2 y_2 - x_1 y_1) + \frac{u_2}{6}(x_2 y_2^3 - x_1 y_1^3) \right] ,
\end{equation}
\begin{equation}
\int_{l_p} x_p dy_p = \frac{1}{2}ab (E_1-E_2) + \frac{1}{2}(x_1y_1-x_2y_2) + \frac{1}{2}x_0(y_1-y_2) - \frac{1}{2}y_0(x_1-x_2) ,
\end{equation}
and
\begin{multline}
\int_{l_p} (3y_p^2+x_p^2)x_p dy_p = \frac{3}{8} ab(a^2+b^2+4x_0^2+4y_0^2)(E_1-E_2)
- \frac{3}{2} ab^2y_0 (\cos{E_1}-\cos{E_2}) \\
- \frac{3}{2} b^2 x_0y_0 (\cos{2E_1}-\cos{2E_2})
- \frac{1}{2} ab^2 y_0 (\cos{3E_1}-\cos{3E_2})
+ \frac{1}{4} b x_0 (9a^2 + 3b^2+4x_0^2+12y_0^2) (\sin{E_1}-\sin{E_2}) \\
+ \frac{1}{4} ab (a^2+3x_0^2+3y_0^2) (\sin{2E_1}-\sin{2E_2})
+ \frac{1}{4} b x_0 (a^2-b^2) (\sin{3E_1}-\sin{3E_2})
+ \frac{1}{32} ab (a^2-3b^2) (\sin{4E_1}-\sin{4E_2}) ,
\end{multline}
respectively. Although the last integral cannot be written in closed form, it can be easily integrated numerically without too many evaluations. For example, for the parameters of Kepler-167e, the numerical integration only requires 30 steps to achieve $<1\,$ppm precision.


\bibliography{arxiv_v2}{}
\bibliographystyle{aasjournal}


\end{document}